\documentclass[twocolumn]{aastex631}

\usepackage{textcomp, gensymb}

\shorttitle{Helium in TOI-1420b}
\shortauthors{Vissapragada et al.}

\begin{document}

\correspondingauthor{Shreyas~Vissapragada}
\email{shreyas.vissapragada@cfa.harvard.edu}

\title{Helium in the Extended Atmosphere of the Warm Super-Puff TOI-1420b}

\author[0000-0003-2527-1475]{Shreyas~Vissapragada}
\altaffiliation{51 Pegasi b Fellow}
\affiliation{Center for Astrophysics $\vert$ Harvard \& Smithsonian, 60 Garden Street, Cambridge, MA 02138, USA}

\author[0000-0002-0371-1647]{Michael~Greklek-McKeon}
\affiliation{Division of Geological and Planetary Sciences, California Institute of Technology, 1200 East California Blvd, Pasadena, CA 91125, USA}

\author[0000-0001-6025-6663]{Dion~Linssen}
\affiliation{Anton Pannekoek Institute for Astronomy, University of Amsterdam, Science Park 904, 1098 XH Amsterdam, The Netherlands}

\author[0000-0002-1417-8024]{Morgan~MacLeod}
\affiliation{Center for Astrophysics $\vert$ Harvard \& Smithsonian, 60 Garden Street, Cambridge, MA 02138, USA}

\author[0000-0002-5113-8558]{Daniel~P.~Thorngren}
\affiliation{Department of Physics \& Astronomy, Johns Hopkins University, Baltimore, MD, USA}

\author[0000-0002-8518-9601]{Peter Gao}
\affiliation{Earth \& Planets Laboratory, Carnegie Institution for Science, Washington, DC, USA}

\author[0000-0002-5375-4725]{Heather~A.~Knutson}
\affiliation{Division of Geological and Planetary Sciences, California Institute of Technology, 1200 East California Blvd, Pasadena, CA 91125, USA}


\author[0000-0001-9911-7388]{David W. Latham}
\affiliation{Center for Astrophysics $\vert$ Harvard \& Smithsonian, 60 Garden Street, Cambridge, MA 02138, USA}

\author[0000-0003-3204-8183]{Mercedes L\'{o}pez-Morales}
\affiliation{Center for Astrophysics $\vert$ Harvard \& Smithsonian, 60 Garden Street, Cambridge, MA 02138, USA}

\author[0000-0002-0371-1647]{Antonija~Oklop{\v{c}}i{\'c}}
\affiliation{Anton Pannekoek Institute for Astronomy, University of Amsterdam, Science Park 904, 1098 XH Amsterdam, The Netherlands}

\author[0000-0001-7144-589X]{Jorge P\'{e}rez Gonz\'{a}lez}
\affiliation{Department of Physics \& Astronomy, University College London, Gower Street, WC1E 6BT London, UK}

\author[0000-0001-9518-9691]{Morgan~Saidel}
\altaffiliation{NSF Graduate Research Fellow}
\affiliation{Division of Geological and Planetary Sciences, California Institute of Technology, 1200 East California Blvd, Pasadena, CA 91125, USA}

\author{Abigail~Tumborang}
\affiliation{School of Physics and Astronomy, University of St. Andrews, North Haugh, St. Andrews, KY16 9SS, UK}

\author[0000-0003-4015-9975]{Stephanie~Yoshida}
\affiliation{Center for Astrophysics $\vert$ Harvard \& Smithsonian, 60 Garden Street, Cambridge, MA 02138, USA}

\begin{abstract}

Super-puffs are planets with exceptionally low densities ($\rho \lesssim 0.1$~g~cm$^{-3}$) and core masses ($M_c \lesssim 5 M_\Earth$). Many lower-mass ($M_p\lesssim10M_\Earth$) super-puffs are expected to be unstable to catastrophic mass loss via photoevaporation and/or boil-off, whereas the larger gravitational potentials of higher-mass ($M_p\gtrsim10M_\Earth$) super-puffs should make them more stable to these processes. We test this expectation by studying atmospheric loss in the warm, higher-mass super-puff TOI-1420b ($M = 25.1M_\Earth$, $R = 11.9R_\Earth$, $\rho = 0.08$~g~cm$^{-3}$, $T_\mathrm{eq} = 960$~K). We observed one full transit and one partial transit of this planet using the metastable helium filter on Palomar/WIRC and found that the helium transits were $0.671\pm0.079\%$ (8.5$\sigma$) deeper than the TESS transits, indicating an outflowing atmosphere. We modeled the excess helium absorption using a self-consistent 1D hydrodynamics code to constrain the thermal structure of the outflow given different assumptions for the stellar XUV spectrum. These calculations then informed a 3D simulation which provided a good match to the observations with a modest planetary mass-loss rate of $10^{10.82}$~g~s$^{-1}$ ($M_p/\dot{M}\approx70$~Gyr). Super-puffs with $M_p\gtrsim10M_\Earth$, like TOI-1420b and WASP-107b, appear perfectly capable of retaining atmospheres over long timescales; therefore, these planets may have formed with the unusually large envelope mass fractions they appear to possess today. Alternatively, tidal circularization could have plausibly heated and inflated these planets, which would bring their envelope mass fractions into better agreement with expectations from core-nucleated accretion. 
\end{abstract}

\section{Introduction} \label{sec:intro}
Thirty years of exoplanet discovery have revealed planetary archetypes for which we have no direct counterparts in our solar system. Among these myriad surprises are the super-puffs, a class of low-mass planets with densities $\rho \lesssim0.1$~g~cm$^{-3}$ \citep{Lee2016, Jontof-Hutter2019}. For planets with $M_p \lesssim 50M_\Earth$, such low densities imply core masses smaller than $5M_\Earth$, which are challenging to reconcile with the core accretion theory for planet formation unless the envelopes are accreted ``dust-free'' \citep{Lee2016} or atop ice-rich cores \citep{Stevenson1984, Venturini2015} beyond the water ice line. Moreover, if super-puffs are indeed as gas-rich as their densities imply, then many should be prone to catastrophic mass loss via photoevaporation and/or boil-off, with atmospheric lifetimes shorter than or comparable to the system ages \citep[i.e. $M_p/\dot{M} \lesssim $~1~Gyr for catastrophic loss;][]{Cubillos2017, Wang2019, Chachan2020}. A number of mechanisms have been proposed to help ease these tensions and bring super-puff envelope mass fractions into closer agreement with expectations from core accretion, including inflation by tides \citep{Millholland2019, Millholland2020} or heat from formation \citep{Masuda2014, Libby-Roberts2020}, planetary rings \citep{Akinsanmi2020, Piro2020, Ohno2022}, and high-altitude aerosols \citep{Kawashima2019, Wang2019, Gao2020, Ohno2021}.

\begin{figure*}[ht!]
    \centering
    \includegraphics[width=\textwidth]{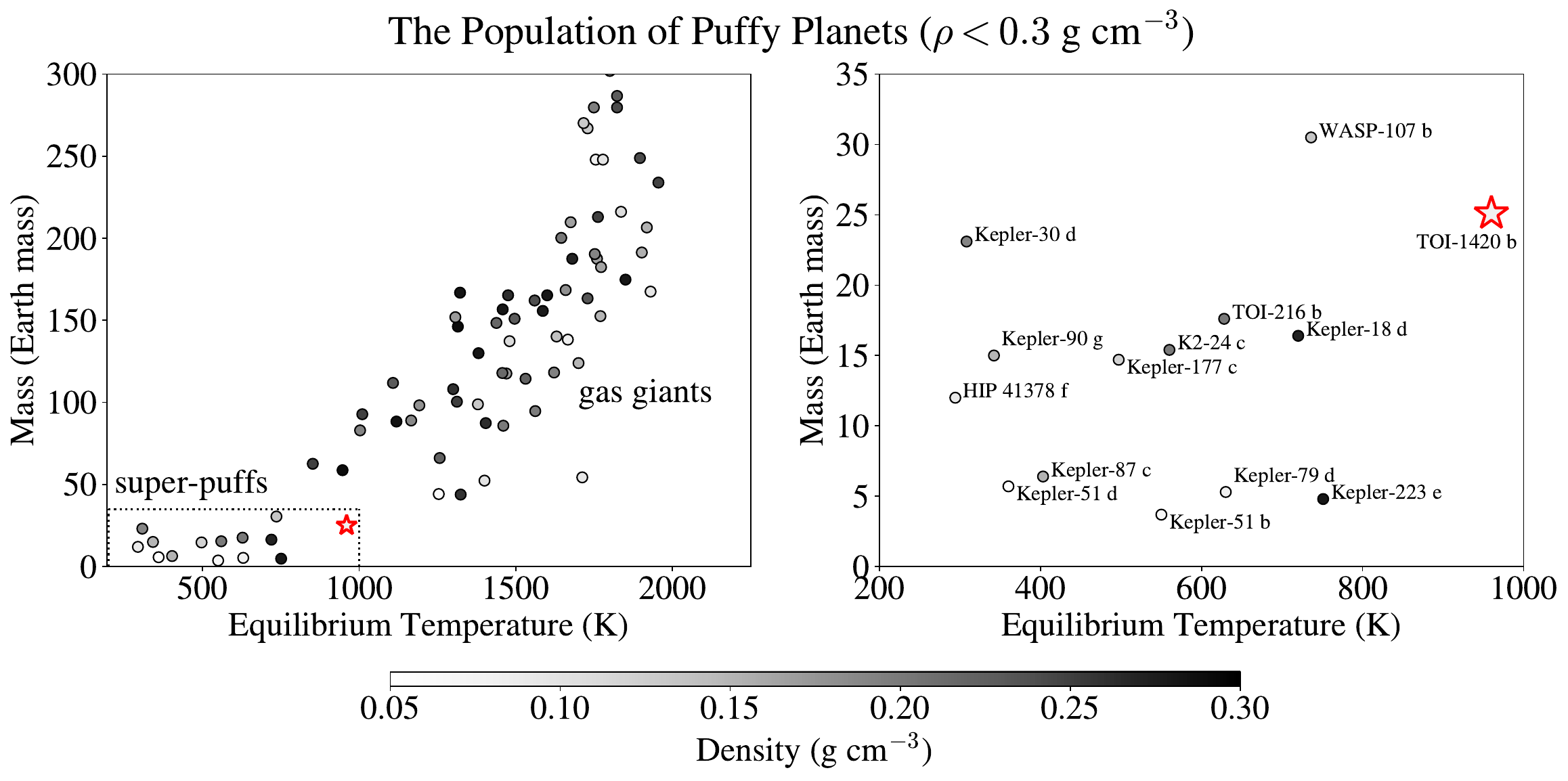}
    \caption{(Left) The masses and equilibrium temperatures of low-density ($\rho < 0.3$~g~cm$^{-3}$) planets on the NASA Exoplanet Archive as of 2023 September 10 \citep{Akeson2013} with masses measured to better than 30\%. Planets on grazing orbits have been removed. TOI-1420b is highlighted with the red star. The shading indicates planetary density, and the dotted lines demarcate the bounds of the plot shown on the right. (Right) Same as left, but focused on the lower-mass super-puff population with individual systems labeled.}
    \label{fig:superpuff}
\end{figure*}

We show the super-puffs in context on a $M_p-T_\mathrm{eq}$ diagram in Figure~\ref{fig:superpuff}. All transiting planets on the NASA Exoplanet Archive \citep{Akeson2013} with densities $\rho < 0.3$~g~cm$^{-3}$ and masses measured to better than 30\% are shown, and planets on grazing orbits have been removed. Most puffy planets hotter than 1000~K are typical gas giants, with large H$_2$/He envelopes that are readily explained by core accretion models \citep[e.g.][]{Lee2019}. Gas giants above this temperature threshold have long been known to have inflated radii driven by the large incident stellar flux \citep[e.g.][]{Bodenheimer2003, Thorngren2018}. Planets usually considered ``super-puffs'' (i.e. those with inferred core masses too small to trigger classical runaway gas accretion) are found in the lower left of this figure. The super-puff and gas giant regimes appear to be separated by a deficit of planets, which may result from sensitivity biases as a function of orbital period: most gas giants have masses measured with radial velocities, and most super-puffs with transit timing variations \citep{Steffen2016, Mills2017}. 

The population is shown in more detail in the right panel of Figure~\ref{fig:superpuff}. We focus on the planets on the upper right corner of the plot, WASP-107b \citep{Anderson2017} and TOI-1420b \citep{Yoshida2023}. Both have inferred core masses $<5M_\Earth$ \citep{Piaulet2021, Yoshida2023}, too small to trigger runaway gas accretion in the traditional core accretion paradigm for gas giants \citep[which requires core masses $\gtrsim 10M_\Earth$; e.g.][]{Bodenheimer1986}. However, these planets stand out as warmer and higher-mass than the other super-puffs. They also differ from the rest of the super-puff population in system architecture: WASP-107b is accompanied only by a distant non-resonant companion \citep{Piaulet2021}, and TOI-1420b also lacks nearby companions \citep{Yoshida2023}. All other super-puffs are found in multi-planet systems near mean-motion resonances, which facilitates their characterization via transit timing variations \citep{Jontof-Hutter2019}, although a few have now also been detected with radial velocities \citep{Petigura2018, Santerne2019}. 

Although their short orbital periods ($P<10$~days) make them more prone to XUV-driven photoevaporation, TOI-1420b and WASP-107b are not expected to be susceptible to catastrophic atmospheric loss, as their gravitational potentials are relatively large despite their low densities \citep{Wang2019, Chachan2020, Belkovski2022}. Indeed, He~1083~nm measurements of WASP-107b suggest an outflow rate of about one Earth mass per gigayear \citep[$M/\dot{M} \approx 30$~Gyr;][]{Spake2018, Allart2019, Kirk2020, Spake2021, Wang2021}, in line with predictions from photoevaporation models \citep{Kubyshkina2018, Caldiroli2022}. This planet has not lost much envelope mass over its lifetime, so it remains plausible that WASP-107b formed with an unusually large envelope \citep{Lee2016, Piaulet2021}. This is a major difference from the lower-mass super-puffs, which would be expected to lose large envelopes on short timescales necessitating many of the aforementioned alternative mechanisms (rings, hazes, etc.). 

In this work, we study atmospheric escape in TOI-1420b \citep{Yoshida2023}. This planet orbits a late G dwarf every 6.96~days, and with $M_p = 25.1\pm3.8M_\Earth$ and $R_p = 11.9\pm0.3R_\Earth$, TOI-1420b has an exceptionally low density of just $0.08\pm0.02$~g~cm$^{-3}$. Moreover, at $J = 10.6$ the system is well-positioned for narrowband photometric observations with the Wide-field InfraRed Camera \citep[WIRC;][]{Wilson2003} on the Hale 200-inch Telescope at Palomar Observatory. We recently used Palomar/WIRC to constrain helium absorption in three systems with similar $J$ magnitudes: HAT-P-18 ($J = 10.8$), WASP-52 ($J = 10.6$), and HAT-P-26  \citep[$J = 10.1$; ][]{Vissapragada2020:helium, Paragas2021, Vissapragada2022:helium}. The former two measurements have recently been confirmed by \textit{JWST} NIRISS/SOSS and Keck/NIRPSEC, demonstrating the utility of narrowband photometry for this science case \citep{Fu2022, Kirk2022}. In Section~\ref{sec:obs}, we describe our narrowband photometric observations of TOI-1420b. We model our narrowband helium light curves in Section~\ref{sec:lightcurves}, interpret the results using outflow models in Section~\ref{sec:massloss}, discuss the implications for warm, high-mass super-puffs in Section~\ref{sec:disc}, and conclude in Section~\ref{sec:conc}.

\section{Helium Light Curves} \label{sec:obs}
We obtained transit photometry of TOI-1420b with Palomar/WIRC. All events were observed with the beam-shaping diffuser \citep{Stefansson2017, Vissapragada2020:ttv}, which spreads the light into a top-hat point-spread function with a 3$\arcsec$ full-width at half-maximum. This allows us to control time-correlated systematics and to integrate for longer on bright sources (improving the duty cycle of the observations). We used the custom metastable helium filter (centered on 1083.3~nm with a bandwidth of 0.64~nm) presented in \citet{Vissapragada2020:helium}.

We observed a full transit of TOI-1420b on UT 2022 June 30 from 05:17 (airmass 1.8) to 12:02 (airmass 1.2) and a partial transit on UT 2022 July 21 from 05:28 (airmass 1.5) to 11:08 (airmass 1.2). The transit duration is $T_{14} = 3.37\pm0.15$~hr \citep{Yoshida2023}. We attempted a partial transit observation on UT 2022 July 28, but poor weather rendered the data unusable for our analysis. All observations were taken with 90~s exposures. To correct the structured background imposed by the narrowband helium filter \citep[further details in][]{Vissapragada2020:helium}, we obtained 8 dither frames immediately before both transit observations. Additionally, we maintained roughly the same detector positioning for both nights, so we observed the same six reference stars each night (which were selected to be the brightest sources in the field besides TOI-1420, ranging from $J = 11.3$ to $J = 8.8$). None of the selected reference stars are known variables, and none exhibited detectable short-term variability in our photometry.

We dark-subtracted, flat-fielded, and background-corrected the data using the methods presented previously in \citet{Vissapragada2020:helium, Vissapragada2020:ttv, Vissapragada2022:helium} and \citet{Greklek-McKeon2023}. We then performed aperture photometry on the target and reference stars using \textsf{photutils} \citep{photutils}. We tested apertures from 5~pixels ($1\farcs25$) to 20~pixels ($5\farcs00$) in radius. The centers of the apertures were allowed to shift to track variations in telescope pointing. Our guiding was stable on both nights, with centroid variations not exceeding 1.5~pixels. We estimated and corrected any residual local background using annular apertures around each source with inner radii of 25~pixels ($6\farcs25$) and outer radii of 50~pixels ($12\farcs50$). The optimal aperture was selected to be the one that minimized the per-point rms scatter in the target star photometry with the average comparison star divided out (removing common-mode systematics) and a moving-median filter applied (removing the transit event). We adopted optimal apertures of 8~pixels ($2\farcs00$) for the first night and 7~pixels ($1\farcs75$) for the second night.

\begin{deluxetable*}{cccccc}[t]
\tablecaption{Priors and posteriors for the light-curve fits. \label{table:posteriors}}
\tablehead{\colhead{Parameter} & \colhead{Unit} & \colhead{Prior} & \colhead{Night 1} & \colhead{Night 2} & \colhead{Night 1 + Night 2}}
\startdata
$P$ & days & $\mathcal{N}(6.9561063, 0.0000017)$ & $6.9561063_{-0.0000013}^{+0.0000013}$ & $6.9561065_{-0.0000013}^{+0.0000013}$ &  $6.9561064_{-0.0000013}^{+0.0000013}$\\
$t_0$ & BJD$_\mathrm{TDB}$ & $\mathcal{N}(2459517.43305, 0.00012)$ & $2459517.433227_{-0.000087}^{+0.000087}$ &  $2459517.433253_{-0.000087}^{+0.000088}$ & $2459517.433238_{-0.000087}^{+0.000087}$\\
$R_\star$ & $R_\Sun$ & $\mathcal{N}(0.923, 0.024)$ & $0.923_{-0.024}^{+0.024}$ & $0.923_{-0.023}^{+0.024}$ & $0.923_{-0.023}^{+0.023}$\\
$a/R_\star$ & -- & $\mathcal{U}(10, 20)$ & $16.37_{-0.36}^{+0.43}$& $16.33_{-0.36}^{+0.42}$ & $16.43_{-0.38}^{+0.44}$\\
$b$ & -- & $\mathcal{U}(0, 1)$ & $0.429_{-0.070}^{+0.047}$ &  $0.429_{-0.072}^{+0.049}$ & $0.414_{-0.077}^{+0.053}$\\
$R_p/R_\star$ (WIRC) & -- & $\mathcal{U}(0, 0.25)$ & $0.1393_{-0.0033}^{+0.0031}$ & $0.1477_{-0.0058}^{+0.0057}$ & 
$0.1409_{-0.0029}^{+0.0028}$\\
$R_p/R_\star$ (TESS) & -- & $\mathcal{U}(0, 0.25)$ & $0.1184_{-0.0017}^{+0.0013}$ & $0.1182_{-0.0018}^{+0.0014}$ &
$0.1179_{-0.0018}^{+0.0014}$\\
$u_1$ (WIRC)  & -- & \citet{exoplanet:kipping13} & $0.36_{-0.23}^{+0.24}$ &   $0.42_{-0.24}^{+0.25}$ & $0.45_{-0.22}^{+0.18}$\\
$u_2$ (WIRC)  & -- &  \citet{exoplanet:kipping13} & $0.09_{-0.28}^{+0.34}$  & $-0.05_{-0.20}^{+0.29}$ & $-0.06_{-0.19}^{+0.31}$\\
$u_1$ (TESS)  & -- &  \citet{exoplanet:kipping13}  & $0.44_{-0.12}^{+0.12}$ & $0.41_{-0.13}^{+0.13}$ & $0.40_{-0.11}^{+0.12}$\\
$u_2$ (TESS)  & -- &  \citet{exoplanet:kipping13}  & $0.09_{-0.23}^{+0.25}$ & $0.15_{-0.25}^{+0.27}$ & $0.17_{-0.24}^{+0.25}$\\
$\delta_\mathrm{WIRC}$ & \% &  --  & $2.176_{-0.086}^{+0.088}$ & $2.43_{-0.19}^{+0.19}$ & $2.228_{-0.080}^{+0.081}$\\
$\delta_\mathrm{comp}$ & \% &  -- & $1.568_{-0.044}^{+0.048}$  & $1.553_{-0.059}^{+0.065}$& $1.556_{-0.040}^{+0.041}$\\
$\delta_\mathrm{mid}$ & \% & -- & $0.608_{-0.085}^{+0.085}$ & $0.88_{-0.18}^{+0.18}$ & $0.671_{-0.079}^{+0.077}$
\enddata
\tablecomments{$\mathcal{N}(a,b)$ indicates a normal distribution with mean $a$ and standard deviation $b$. $\mathcal{U}(a,b)$ indicates a uniform distribution between $a$ and $b$. The WIRC depth $\delta_\mathrm{WIRC}$, comparison depth $\delta_\mathrm{comp}$, and mid-transit excess depth $\delta_\mathrm{mid} = \delta_\mathrm{WIRC} - \delta_\mathrm{comp}$ are all derived parameters. Posteriors for the detrending weights, WIRC jitter parameters, and TESS error scaling parameter are included in Table~\ref{table:extraposteriors}}.
\end{deluxetable*}

 \begin{figure*}[ht!]
    \centering
    \includegraphics[width=\textwidth]{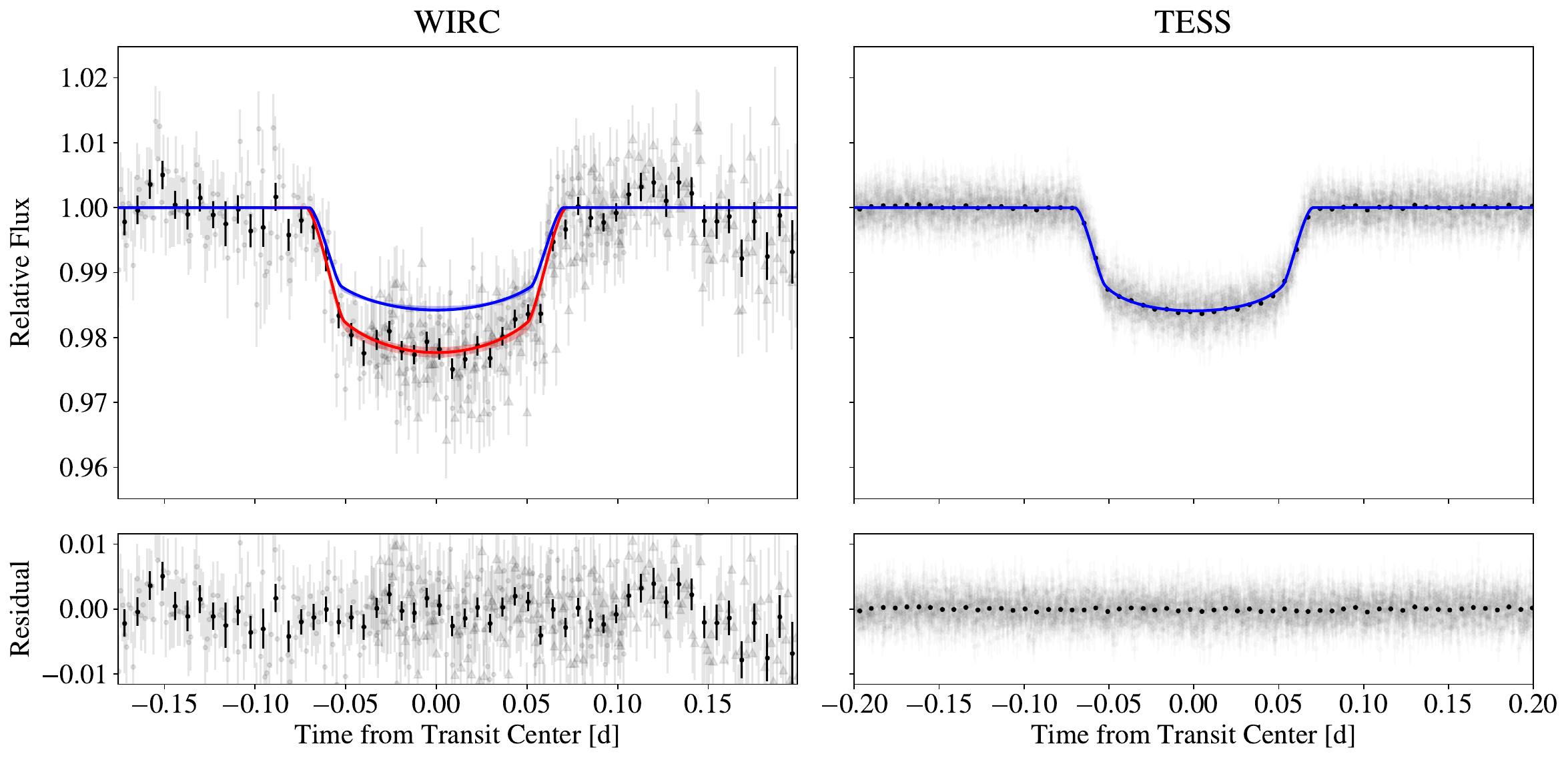}
    \caption{WIRC (left) and phase-folded TESS (right) light curves of TOI-1420b. The data are shown unbinned in gray (with circles and triangles denoting points from the first and second nights of WIRC observations, respectively) and binned to 10~min cadence in black. Maximum a posteriori (MAP) light curve models are overplotted in the solid curves for WIRC (red) and TESS (blue), with the shaded regions denoting the 68\% credible intervals. The TESS light curve model is recalculated using the helium bandpass limb darkening coefficients on the WIRC plot, showing that differences in limb darkening cannot explain the increased depth at mid-transit.}
    \label{fig:sidebyside}
\end{figure*}

\section{Light Curve Modeling} \label{sec:lightcurves}

We first fit each helium curve separately using the \textsf{exoplanet} code \citep{exoplanet:joss, exoplanet:zenodo}. We fit the target light curve $f$ as the product of a transit light curve model $T$ and a systematics model $S$:
\begin{equation}
    f = T(R_\mathrm{p}/R_\star, a/R_\star, b, u_1, u_2, P, t_0) \times S.
\end{equation}
To model the transit, we included a uniform prior on the planet-to-star radius ratio $R_\mathrm{p}/R_\star$, scaled semimajor axis $a/R_\star$, and impact parameter $b$, uninformative priors \citep{exoplanet:kipping13} on the quadratic limb darkening coefficients $(u_1, u_2)$, and normal priors on the planetary orbital period $P$, transit epoch $t_0$, and stellar radius $R_\star$. Normal priors were taken from \citet{Yoshida2023}, and are reproduced in Table~\ref{table:posteriors}. The planet does not appear to be on an eccentric orbit \citep[$e < 0.17$][]{Yoshida2023} so we fixed the eccentricity to 0. We tracked the depth of the WIRC light curve at mid-transit $\delta_\mathrm{WIRC}$ numerically rather than using $(R_p/R_\star)^2$ so that limb darkening was taken into account.

We included the two~minute cadence TESS light curve of TOI-1420b from \citet{Yoshida2023} in all our fits to provide a reference $R_p/R_\star$ for the WIRC light curves. The TESS data also helped us to constrain the degeneracy between $b$ and $a/R_\star$ with the precise transit shape. All transit parameters were shared between the WIRC and TESS fits except for $R_p/R_\star$ and the quadratic limb darkening coefficients $(u_1, u_2)$. When we tracked the comparison transit depth $\delta_\mathrm{comp}$, we recalculated the TESS light curve with the WIRC limb darkening coefficients. This excluded the possibility that perceived depth differences were caused by differences in limb darkening between the two bandpasses. The excess depth at mid-transit, which describes the extra absorption due to metastable helium, was then calculated as $\delta_\mathrm{mid} = \delta_\mathrm{WIRC} - \delta_\mathrm{comp}$. 

We modeled the systematics $S$ as a linear combination of detrending vectors $s_i$ including the six comparison star light curves and the mean-subtracted times (defining a linear trend in time):
\begin{equation}
    S = \sum_i w_i s_i
\end{equation}
The weight $w_i$ for each detrending vector $s_i$ in the linear combination was a free parameter with uniform prior $\mathcal{U}(-2, 2)$. We also tested including the airmass, the centroid offsets, and the water absorption proxy described previously in \citet{Paragas2021} and \citet{Vissapragada2022:helium} as additional detrending vectors, and calculated the Bayesian Information Criterion \citep[BIC;][]{Schwarz1978} to determine whether the increase in log-likelihood justified the addition of these parameters to the model. However, none of these additional detrending vectors improved the BIC, i.e. the marginal increase in final log-likelihood did not justify the addition of any of the parameters, so we left these out of the final models. Finally, we included a photometric jitter term describing the excess white noise in the WIRC data (added in quadrature to the photon noise for each point) and a multiplicative scaling factor for the uncertainties on the TESS data.

In all of the fits, we first optimized towards the maximum a posteriori (MAP) solution with \textsf{pymc3} \citep{exoplanet:pymc3}. We identified and removed 5$\sigma$ outliers from the residuals to the MAP fit, iterating the MAP fit and outlier rejection procedure until a stable MAP solution was obtained. Then, we used the No U-Turn Sampler \citep{Hoffman2011} in \textsf{pymc3} to sample the posterior distributions for the model parameters. Every fit was run with 10,000 tuning steps before taking 10,000 posterior draws (distributed across two chains). To assess convergence, we verified that the Gelman-Rubin statistic \citep{Gelman1992} $\hat{R} \ll 1.01$ for all sampled parameters, and we also visually inspected the chains to ensure they were well-mixed. The posteriors of our single-planet fits are given in Table~\ref{table:posteriors}.

The transit depths on the two nights were consistent at about the 1$\sigma$ level. The WIRC transit depth was less precisely constrained on the second night than the first because of the partial phase coverage, but the second night also offered more post-egress baseline than the first night. Given the reasonable agreement in $R_p/R_\star$ between both datasets and the complimentary baseline coverage, we decided to fit both WIRC light curves jointly with the TESS light curve. The posteriors of the joint fit are included in Table~\ref{table:posteriors}. Figure~\ref{fig:sidebyside} shows the joint WIRC light curve alongside the phase-folded TESS data. The final WIRC transit was $0.671\pm0.079\%$ deeper than the TESS transit, indicating substantial extra absorption by metastable helium in the upper atmosphere of TOI-1420b at a confidence of 8.5$\sigma$. 

\begin{figure}[ht!]
    \centering
    \includegraphics[width=0.35\textwidth]{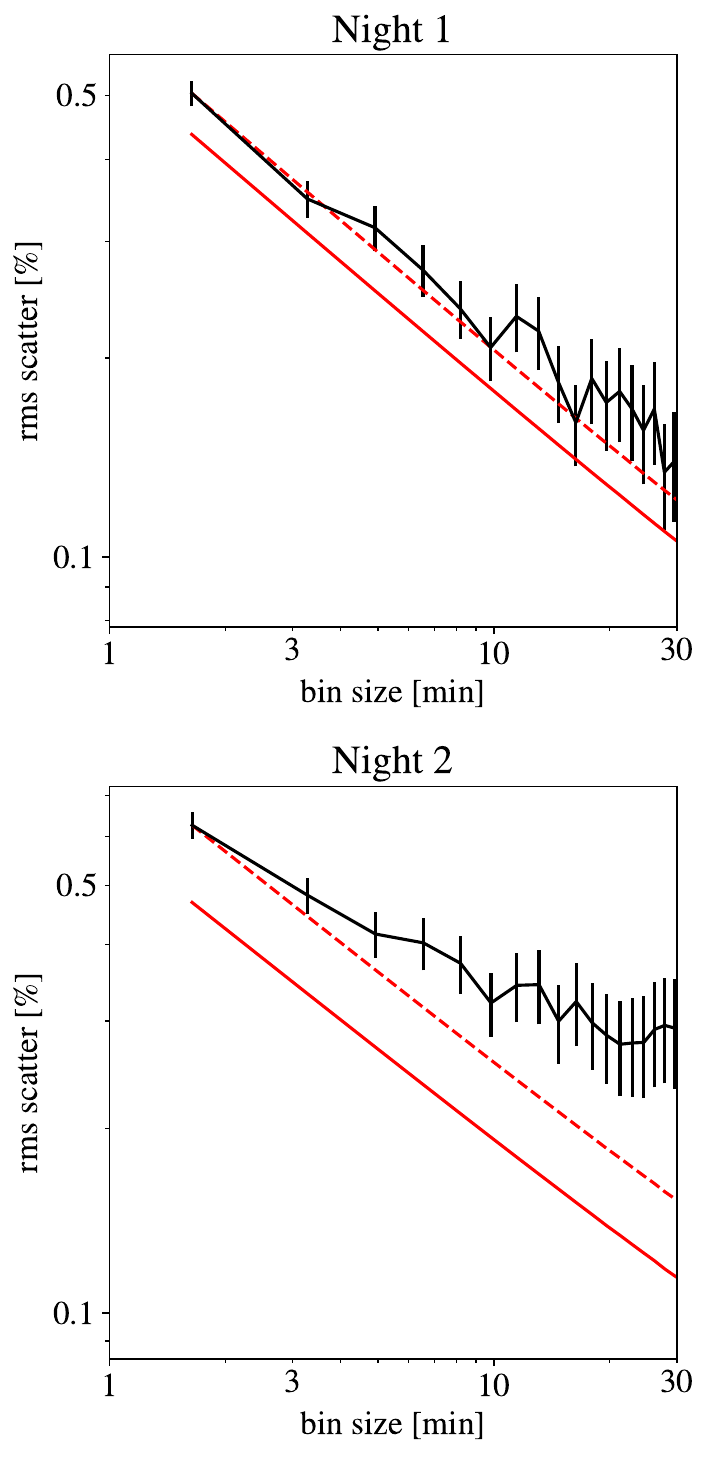}
    \caption{Allan deviation plots for the two WIRC light curves. The black curves indicate the rms scatter as a function of bin size on each night, the solid red curves indicate expectations from photon noise, and the dashed red curves are the solid red curves scaled up to match the first point of the black curve, i.e. the expectation if our data binned down exactly like white noise.}
    \label{fig:allan}
\end{figure}

The rms scatter of the unbinned data was 0.50\% on the first night and 0.63\% on the second night. These were $1.2\times$ and $1.3\times$ the photon noise on each night, respectively, indicating good photometric performance. This is the closest to photon noise we have been able to achieve with Palomar/WIRC using the helium filter \citep{Vissapragada2022:helium}, likely because six good comparison stars were available in the field. However, when reducing the second night of data we noticed a large correlated noise component at about a 30~min timescale in the Allan deviation plots (Figure~\ref{fig:allan}). A correlated feature with this timescale is clearly visible at the end of the WIRC light curve in Figure~\ref{fig:sidebyside}, though the uncertainties in this region of the light curve suggest it is not very significant. This feature was not clearly associated with rapid changes in water absorption proxy, centroid offset, or total flux. Additionally, the second night did not contribute much information to the final excess depth measurement ($\delta_\mathrm{mid} = 0.608^{+0.085}_{-0.085}\%$ for the first night only and $\delta_\mathrm{mid} = 0.671^{+0.077}_{-0.079}\%$ for the joint fit), so we proceeded without treating this feature in detail.

\section{Mass-Loss Modeling} \label{sec:massloss}


We approached modeling TOI-1420b's helium absorption signal in three stages. First, we estimated the characteristic outflow sound speeds and temperatures that correspond to such a low-density planet. Next, we calculated one-dimensional models treating the thermodynamics self-consistently rather than assuming an isothermal wind. Finally, informed by the results of the 1D modeling, we calculated a 3D model which treats the thermodynamics approximately, but captures the interaction of the outflow with the tidal and stellar wind environment.

\subsection{Characteristic temperatures}

TOI-1420b's low density carries some important implications for the modeling of its outflow. The one-dimensional, transonic Parker wind models often adopted for modeling metastable helium absorption \citep[e.g.][]{Oklopcic2018, Lampon2020, DosSantos2022:pwinds} begin with subsonic flow near the planet that accelerates to supersonic flow at larger radii. A transonic wind requires $R_s > R_p$, where: 
\begin{equation}
    R_s \approx \frac{GM_p}{2c_s^2},
\end{equation}
\citep[the tidal gravity correction is small for TOI-1420b; see Appendix E of][]{Vissapragada2022:helium}. The sound speed of the isothermal outflow is:
\begin{equation}
    c_s = \sqrt{\frac{k_\mathrm{B}T_0}{\bar{\mu}m_\mathrm{H}}},
\end{equation} where $k_\mathrm{B}$ is Boltzmann's constant, $T_0$ is the isothermal sound speed, $\bar{\mu}$ is the mean molecular weight and $m_\mathrm{H}$ is the mass of a hydrogen atom. The condition $R_s > R_p$ can equivalently be written $\Phi > 2c_s^2$, where $\Phi = GM_p/R_p$ is the gravitational potential. 

TOI-1420b has a small gravitational potential of $\Phi \approx 10^{12}~$erg~g$^{-1}$, so only isothermal outflows with $c_s \lesssim 8$~km~s$^{-1}$ yield transonic velocity structures (i.e. $R_s > R_p$). In terms of an isothermal temperature, this corresponds to $T_0\lesssim 7700$~K for an assumed mean molecular weight of $\mu =1$~amu. This condition overlaps significantly with the range of isothermal outflow temperatures usually assumed when modeling outflows from hot giant planets \citep[$5000\mathrm{K}\lesssim T_0\lesssim15000\mathrm{K}$; e.g.][]{Linssen2022, Vissapragada2022:efficiency}. To avoid the issue of applying the isothermal Parker wind outside the bounds of its validity, we decided to consider one-dimensional models with self-consistent (non-isothermal) thermodynamics and the nature of the transonic ouflows that develop. 

\subsection{One-dimensional outflow models}

\begin{figure}
    \centering
    \includegraphics[width=0.5\textwidth]{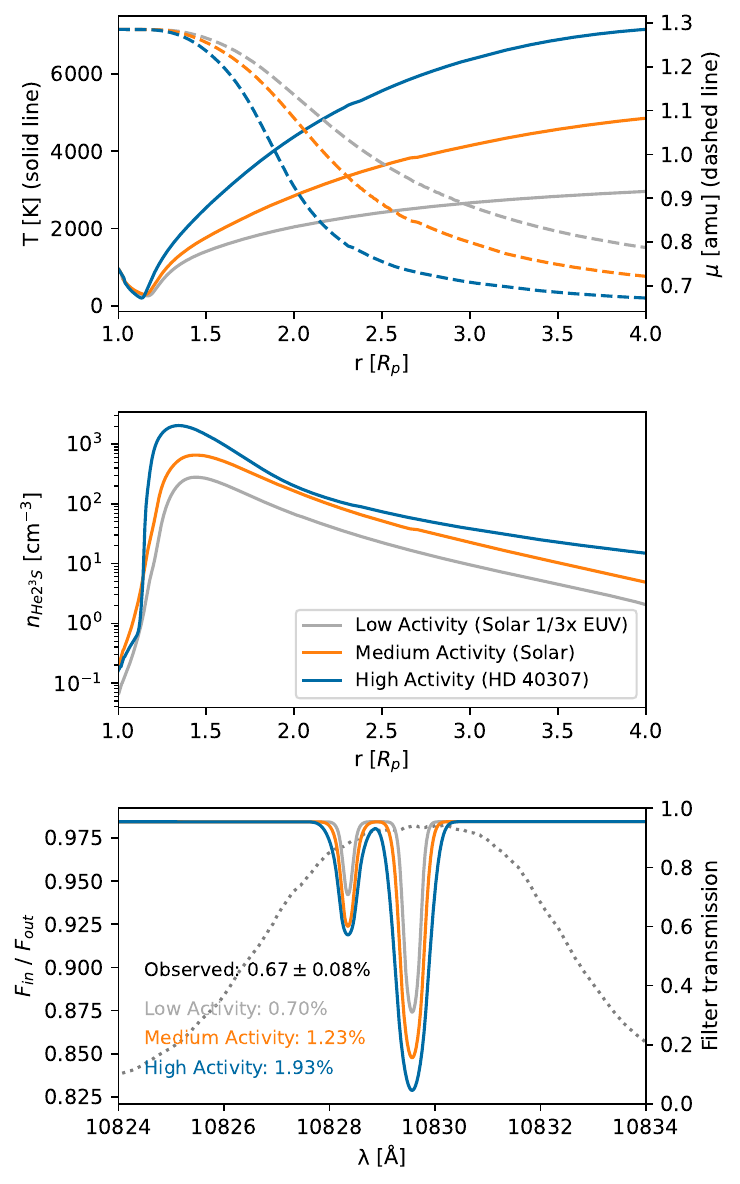}
    \caption{Temperature (top panel, solid curves), mean molecular weight (top panel, dashed curves), and metastable helium number density (middle panel) profiles for the high activity (blue), medium activity (red), and low activity (green) \textsf{ATES+Cloudy} models. The solid curves in the bottom panel show the predicted metastable helium features at high dispersion in air wavelengths in the observatory rest frame. The dotted curve shows the filter transmission function. The text annotations give the corresponding excess transit depth values (after integrating through the filter transmission function) compared to the observed absorption.}
    \label{fig:dionfigure}
\end{figure}

To self-consistently treat the wind-launching physics, we used the 1D photoionization hydrodynamics code \textsf{ATES} to compute outflow density, velocity, and temperature profiles \citep{Caldiroli2021}. Contrary to an isothermal Parker wind profile, \textsf{ATES} calculates the temperature structure based on the stellar irradiation assuming a pure H/He composition for the wind. With the equilibrium temperature as a boundary condition at $R_p$, the wind starts hydrostatically and is indeed transonic. We then passed the outflow structure to the NLTE photoionization code \textsf{Cloudy} \citep{Ferland1998, Ferland2017} to obtain the metastable helium density profile. We finally took \textsf{Cloudy}'s metastable helium density profiles and used a custom radiative transfer module \citep[based on][]{Oklopcic2018, Linssen2022} to calculate the mid-transit spectrum in the stellar rest frame, using the limb-darkening coefficients and transit impact parameter of Table~\ref{table:posteriors}. We assumed a hydrogen/helium abundance ratio of 90/10 by number. Finally, we Doppler shift the spectrum from the stellar rest frame into the observatory rest frame, taking into account the system's barycentric radial velocity of $-10.33$~km~s$^{-1}$ \citep{Yoshida2023} as well as the Earth's orbital motion relative to the Solar System barycenter on the nights of observation. Finally the excess absorption in the WIRC filter passband is calculated following Equation (3) of \citet{Vissapragada2020:helium}.

The XUV flux of the host star is an important ingredient in the simulations, as it governs both the thermal structure in \textsf{ATES} and the metastable helium population in \textsf{Cloudy}. However, the high-energy spectrum of TOI-1420 is unknown (and challenging to measure at $d = 200$~pc), so we used spectra of similarly active stars as proxies. TOI-1420 is a late G dwarf ($T_\mathrm{eff} = 5493$~K) with $\log(R'_{HK}) = -4.87$ \citep{Yoshida2023}, indicating similar chromospheric activity to the mean solar value of $-4.91$ \citep{Mamajek2008}. To cover the uncertainty in the stellar XUV spectrum, we repeated the modeling using three different proxies: the MUSCLES spectrum of HD 40307 \citep[an early K star representing ``high activity'' albeit with a modest $\log(R'_{HK}) = -4.99$;][]{Mayor2009, France2016, Loyd2016, Youngblood2016}, a quiet solar spectrum \citep[representing ``medium activity'';][]{Oklopcic2018, DosSantos2022:pwinds}, and the solar spectrum with EUV ($<1000$~\AA) scaled down by a factor of three (representing ``low activity''). 

The high and medium activity models predict mass-loss rates of $10^{11.25}$~g~s$^{-1}$ and $10^{10.93}$~g~s$^{-1}$, respectively, but both predict much larger helium signals than we observed (Figure~\ref{fig:dionfigure}). The low activity spectrum gives the best match to the data at a mass-loss rate of $10^{10.58}$~g~s$^{-1}$, and we therefore adopt this as our preferred XUV model for the star. At the same time, the high and medium activity solutions should not be thought of as ``ruled out'' on the basis of the 1D modeling alone. Some of the absorption in these models comes from high altitudes $>4R_p$ where shadowing, tidal effects, and interaction with the stellar wind becomes important for predicting the metastable helium signature \citep{Wang2021, MacLeod2022}. These effects are not treated in ATES, so rather than reducing the activity in the 1D models further, we turned to 3D modeling informed by the results of the self-consistent 1D analysis.

\subsection{Three-dimensional outflow model} \label{3d}

In order to assess the impact of the stellar environment on the outflow of TOI-1420b -- including the stellar gravity and stellar wind -- we perform a three-dimensional hydrodynamic calculation of the escape of a planetary outflow similar to that of the TOI-1420 system. Our model is based on the \textsf{Athena++} Eulerian hydrodynamic code \citep{Stone2008, Stone2020} and the planetary outflow modeling and radiative transfer post processing software developed by \citet{MacLeod2022}. This simulation does not include magnetic fields, which can suppress the planetary helium signature \citep{Schreyer2023}.

We simulate the interaction of planetary and stellar wind outflows in the corotating reference frame. We assume that the planet and star rotate with the orbital frequency, and adopt the basic system parameters of Table \ref{table:posteriors}. In our model the planetary and stellar winds are simulated with a nearly-isothermal adiabatic index of $\gamma=1.0001$, but have different characteristic temperatures. The planetary outflow is launched from a uniform boundary condition representing the planet, and is defined by the hydrodynamic escape parameter $\lambda_{\rm p} = GM_p/R_p c_s^2$, where $c_s$ is the sound speed of the isothermal planetary outflow. We adopt $\lambda_p = 5$, which corresponds to $T\approx 3200$~K for $\mu= 1$~amu. The value of the hydrodynamic escape parameter is chosen to roughly correspond to the 1D ATES model with the $1/3\times$~Solar EUV spectral energy distribution in Figure~\ref{fig:dionfigure}. Similarly, on the basis of the ATES modeling, we adopt the $1/3\times$~Solar EUV as our spectrum for the three-dimensional model spectra.

With these basic parameters of the system set by either the known system properties (e.g. Table \ref{table:posteriors}), or by comparison to the 1D ATES models, we are free to vary the relative strength of the planetary and stellar outflows in our hydrodynamic calculation by specifying the stellar and planetary boundary conditions that generate these flows. This freedom allows us to scale our calculation to reproduce the mid-transit excess absorption. We chose the density of our planetary boundary condition such that the measured emergent outflow rate in steady state is $6.7\times 10^{10} $~g~s$^{-1}$ ($10^{10.82}$~g~s$^{-1}$). The stellar boundary condition is set similarly, but with a hydrodynamic escape parameter of $\lambda_\ast = 10$, and a mass-loss rate of $1.6\times 10^{11} $~g~s$^{-1}$, approximately $3.2\times$ the planetary mass loss rate, or $\approx 3.5\times 10^{-15} M_\odot$~yr$^{-1}$. We note that this last choice is somewhat arbitrary, as only much (e.g. order-of-magnitude) stronger or weaker stellar winds are ruled out. Finally, we evolve the system for roughly 6 orbital periods (41.7~d), and find that the flow reaches a steady-state after $\sim 10$~d. 

\begin{figure}
    \centering
    \includegraphics[width=0.5\textwidth]{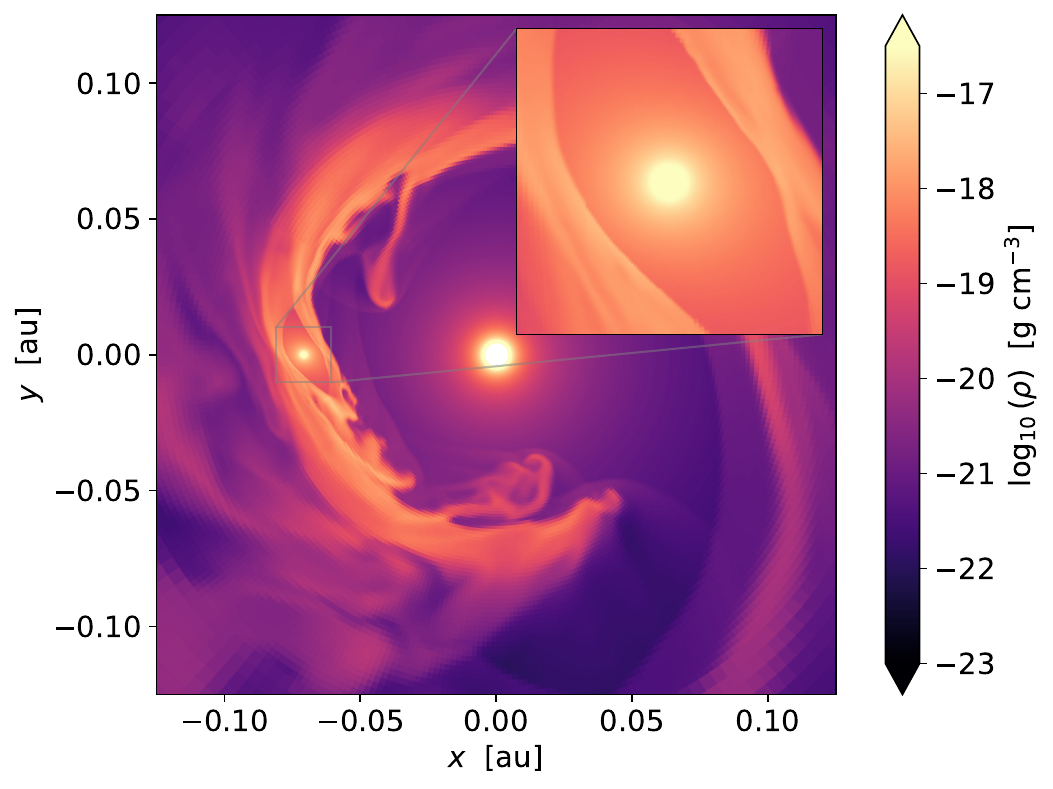}
    \includegraphics[width=0.5\textwidth]{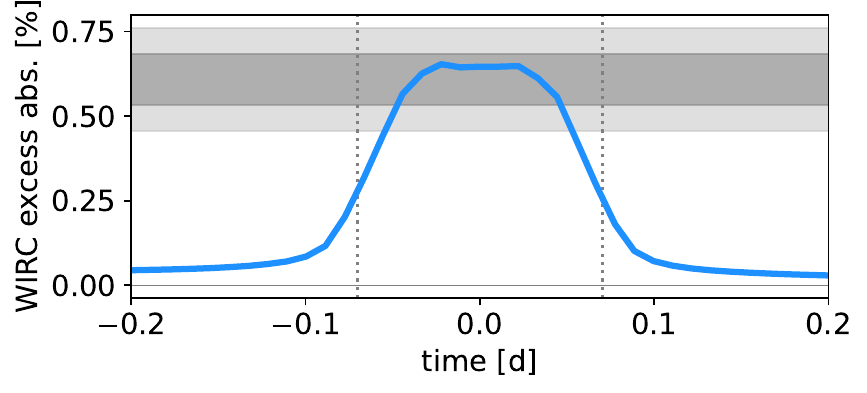}
    \caption{Large-scale model of the planetary outflow from TOI-1420b. The upper panel shows the logarithm of the mass density $\log_{10}(\rho)$ of a slice through the orbital plane of the three-dimensional hydrodynamic model. A small bubble of unshocked mass loss around the planet is truncated by the star's tidal gravitational field, after which the outflow extends into leading and trailing tails ahead and behind the planet in its orbital path, respectively. The lower panel shows the corresponding excess absorption in the WIRC helium bandpass as a function of time during a transit, and shaded regions denote the 1$\sigma$ (dark gray) and 2$\sigma$ (light gray) uncertainties on the measured mid-transit excess absorption. Despite the extended tails, most of the absorption comes during the optical transit (approximately $\pm 0.07$~d, indicated by vertical lines) with a gradual return to very low levels of excess absorption outside of $\pm0.1$~d. This indicates that most of the optical depth contributing to the excess absorption lies in dense regions at the base of the planetary outflow near the planet -- not from the large-scale structure, which is mostly transparent.}
    \label{fig:morganfig}
\end{figure}

The upper panel of Figure \ref{fig:morganfig} shows the emergent flow in a slice through the orbital plane. The star lies at the origin, and the planet is located in the $-x$-direction. An inset shows the flow in a region of $\pm20R_p$ (roughly $\pm0.1$~au). As the planetary outflow expands into the stellar environment, it is shaped by the tidal gravity and stellar wind. Near the planet, outflowing material is initially roughly homogeneous. Outside the planetary Hill sphere of radius $\approx 4.1 R_p$, however, the stellar gravity dominates and the nearly-radial outflow from the planet is truncated by shocks and sheared into tidal tails leading and trailing the planet. Interaction with the stellar wind also partially shapes these tails and their kinematics in our model, though we note that the value of the stellar wind mass loss rate is not measured in TOI-1420, so the influence of the stellar wind could be larger or smaller than our nominal assumption \citep[e.g. as described by][]{MacLeod2022}. 

We post-process metastable helium level populations and transit spectra in the snapshot shown in Figure \ref{fig:morganfig}. For the stellar SED, we adopt the low-activity model described in the preceding section, with 1/3 of the solar EUV flux. We adopt an observer in the general $-x$-direction but at different viewing angles in order to represent the passage of time as the planet transits. The mid-transit model spectrum is similar to the low-activity model of the preceding section. We then compute the excess absorption in the WIRC filter passband following Equation~(3) of \citet{Vissapragada2020:helium}.  

The resulting excess absorption light curve is shown in the lower panel of Figure \ref{fig:morganfig}. This model achieves similar excess depth at mid-transit to that observed in Figure~\ref{fig:sidebyside}, and shows that the vast majority of absorption is concentrated during the optical transit duration. The extended tails of planetary atmosphere material are effectively hidden by their low optical depth. Thus the region that forms the metastable helium absorption lies within a few planetary radii -- near the peak of metastable helium density in the center panel of Figure \ref{fig:dionfigure}. At these scales the planetary outflow is nearly spherical and our models thus predict spectra similar to those of the 1D models despite the complexity of the large-scale flow. This optical depth effect can be contrasted with outflows like those observed for HAT-P-32b and HAT-P-67b, where the larger mass-loss rates give the tails greater optical depths, making them readily observable even when compared to the mid-transit absorption \citep{Gully-Santiago2023, Zhang2023}. 

\section{Discussion} 
\label{sec:disc}

\subsection{Can TOI-1420b retain its puffy hydrogen-rich atmosphere?}
Atmospheric retention has been a major puzzle for super-puff planets. Assuming their measured radii correspond to typical photospheric pressures on the order of 10~mbar, many lower-mass super-puffs with $M_p \lesssim 10M_\Earth$ (including e.g. Kepler-51b, Kepler-51d, Kepler-79d, Kepler-87c, and Kepler-223e) are expected to be unstable to catastrophic atmospheric escape via photoevaporation or boil-off. This has led to suggestions that the measured $R_p$ must correspond to lower pressures in these planets, which could be a result of high-altitude clouds, dust, hazes, or rings \citep{Lammer2016, Cubillos2017, Wang2019, Chachan2020, Libby-Roberts2020, Piro2020, Ohno2022}. On the other hand, higher-mass super-puffs with $M_p \gtrsim 10M_\Earth$ (including e.g. TOI-1420b, WASP-107b, Kepler-177c, and HIP 41378f) are predicted to be stable against catastrophic mass loss because of their (relatively) large gravitational potentials \citep{Kubyshkina2018, Chachan2020, Belkovski2022, Caldiroli2022}. 

Our helium observations of TOI-1420b agree well with the theoretical prediction that high-mass super-puffs should be stable against atmospheric loss. The three-dimensional outflow model in Section~\ref{sec:massloss} matched the observed helium signature at a mass-loss rate of $\dot{M} = 10^{10.82}$~g~s$^{-1}$, corresponding to a mass-loss timescale $M/\dot{M} \approx 100$~Gyr. The situation is similar for this planet's doppelg\"{a}nger WASP-107b, for which \citet{Wang2021} similarly found a good fit to helium observations with $M/\dot{M} \approx 30$~Gyr; moreover their model provided a remarkably good match to the observed post-transit tail for WASP-107b \citep{Spake2021}. The inferred outflow rates for both these planets are not expected to result in significant mass loss over the planetary lifetime. Therefore, despite their remarkably low densities, super-puffs with $M_p \gtrsim 10M_\Earth$ indeed appear stable against atmospheric loss. 

\subsection{Is TOI-1420b tidally-inflated?} \label{sec:tides}
While atmospheric retention may not be a problem for high-mass super-puffs, it remains challenging to form planets with such large envelope mass fractions \cite[$\gtrsim80\%$][]{Piaulet2021, Yoshida2023} within an in situ core accretion framework \citep{Lee2019}. One possible resolution is that these planets accreted their envelopes far from their stars ($\gtrsim1$~au), where dust-free atmospheres could cool and accrete rapidly \citep{Lee2016, Piaulet2021}. Migration would then be required to bring the planets to their observed positions, and indeed many super-puffs are observed to be parts of resonant chains that suggest such a migration history \citep{Jontof-Hutter2019}. However, TOI-1420b and its doppelg\"{a}nger WASP-107b do not reside in resonant configurations. WASP-107b is accompanied by a distant companion that might have driven past migration onto the current high-obliquity orbit \citep{Dai2017, Piaulet2021, Rubenzahl2021}, but no additional companions are yet known in the TOI-1420 system \citep{Yoshida2023}.

We therefore turn to tidal inflation as a possible alternative explanation for the low apparent density of TOI-1420b. Although tidal inflation is not required to match the outflow data, the \textsf{ATES}+\textsf{Cloudy} approach used in Section~\ref{sec:massloss} is agnostic to this mechanism. The underlying envelope mass fraction is immaterial to the model so long as the density and temperature at the lower boundary ($R_p$) are correctly described. Moreover, the overall escape rate of the flow is relatively insensitive to the boundary density \citep{Salz2016, Caldiroli2022}. For WASP-107b, the radial velocity data permit a small eccentricity ($e < 0.14$ at 2$\sigma$) and the resulting heating from eccentricity tides could bring the true envelope mass fraction below 50\% \citep{Millholland2020, Piaulet2021}, though this conclusion depends sensitively on the unknown planetary obliquity and tidal quality factor. The radial velocity curve for TOI-1420b permits a similarly small eccentricity \citep[$e < 0.17$ at 2$\sigma$; ][]{Yoshida2023}, so the same mechanism may apply. 

\begin{figure}[ht!]
    \centering
    \includegraphics[width=0.5\textwidth]{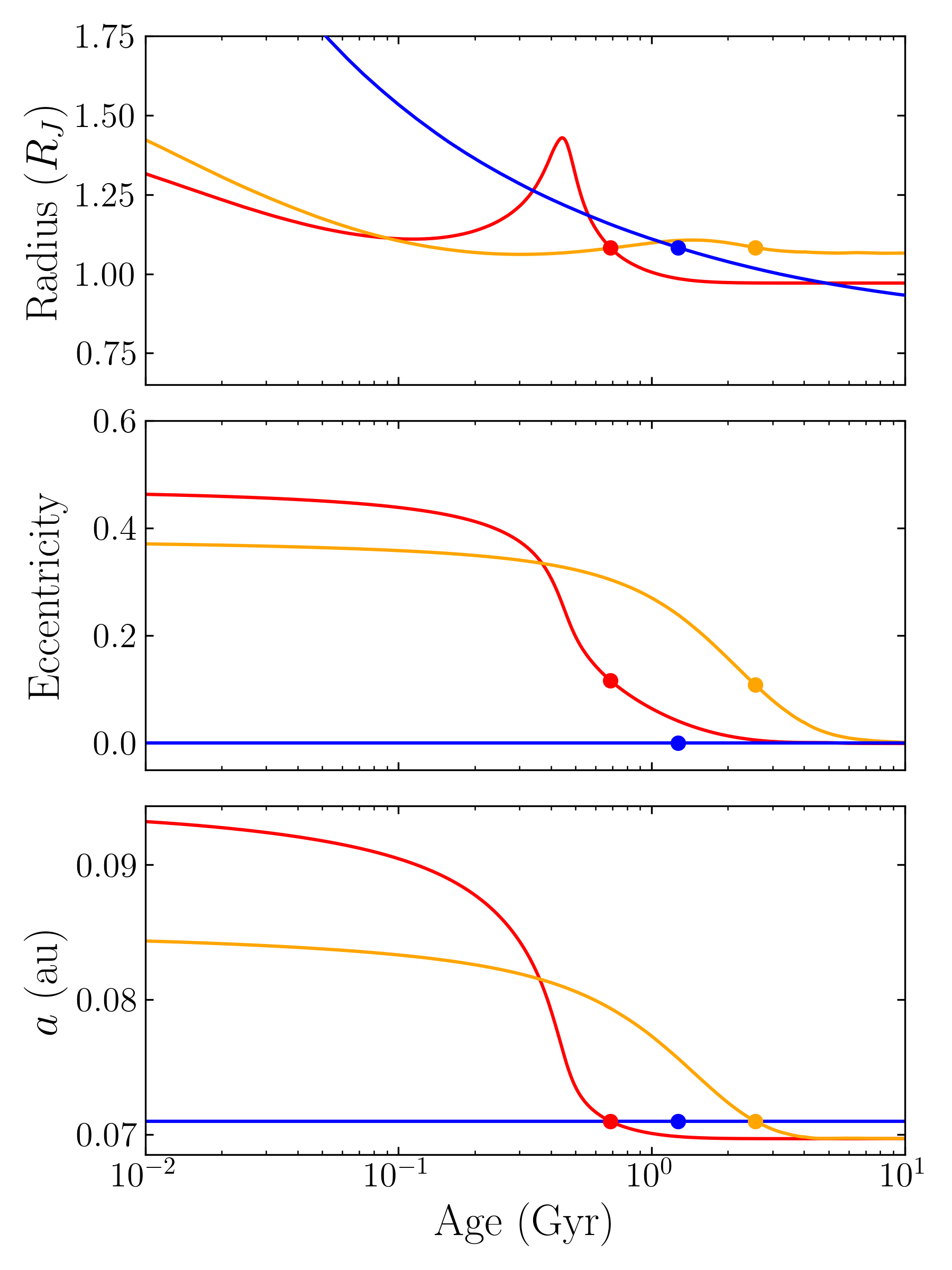}
    \caption{The radius, eccentricity, and semimajor axis evolution for several possible cases of TOI-1420b's tidal evolution. The colored points indicate parts of each evolution track that match the planet's present-day radius ($1.061\pm0.029R_\mathrm{J}$), eccentricity ($<0.17$), and semi-major axis ($0.0710\pm0.0012$~au). The blue curves indicate a planet that began on a circular orbit, where the large radius is due entirely to a low bulk metallicity (in this example, $Z_p=0.13$). In red is the "young circularizing" case, in which the planet is relatively young ($\sim1$~Gyr) and at the final stages of fairly rapid circularization ($\log(Q_p)=5.5$), allowing for a hotter planet with larger bulk metallicity $Z_p=0.37$. Finally, the yellow curves indicate a slower circularization track ($\log(Q_p)=6$, $Z_p=0.33$), which is a potential compromise between the young age of the red line and the very low metallicity blue line. These are only three illustrative examples of a larger solution space (see Section~\ref{sec:tides}).}
    \label{fig:tides}
\end{figure}

To investigate the possibility that tidal circularization has heated and inflated the planet, we constructed a model which combines the giant planet thermal evolution models of \citet{Thorngren2018} with the equilibrium tides model \citep[e.g.][]{Goldreich1966, Mardling2002, Jackson2008}. Similarly to \citet{Lopez2016}, we are not including tidal inspiral or obliquity tides in this model, but we do calculate the amount of heating resulting from circularization to add to the energy balance of the interior. Thus for a constant planetary $M_p$, bulk metallicity $Z_p$, and tidal quality factor $Q_p$, we evolve the eccentricity, semimajor axis, and interior specific entropy over time. 

There is a broad space of possible solutions for TOI-1420b, as the model has a number of free parameters ($e_0$, $a_0$, $Q_p$, $Z_p$ and age) but only strong constraints on $a$ and $R_p$. Therefore, we provide three illustrative evolution tracks for TOI-1420b in Figure~\ref{fig:tides}. For reference we show the case in which the planet formed on a circular orbit, which requires an unusually low metallicity of $Z_p=0.13$ for a planet of this mass to match the radius \citep[for more details on this nominal model, see Section~4 of][]{Yoshida2023}. This model corresponds to a maximum core mass (assuming all metals are confined to a rocky core) of $Z_pM_p = 3.3M_\Earth$. If instead the planet started with an eccentric orbit, the semi-major axes and eccentricity follow a sigmoid curve with the logarithm of the age. During this process of circularization a substantial portion of the orbital energy is injected into the planet, resulting in an increase in the planet's radius which decays as the excess heat is radiated away. The red and yellow curves of Figure~\ref{fig:tides} show how we may be observing the planet towards the tail end of this process, after it has lost most of its initial eccentricity \citep[consistent with the upper limit from][]{Yoshida2023} but while it still has a moderately hotter interior than it would otherwise. This extra heat allows us to reproduce the observed radius with higher bulk metallicities $Z_p = 0.3-0.4$, implying core masses that are more in line with core-nucleated accretion \citep[similarly to the case for a tidally-heated WASP-107b;][]{Millholland2020, Piaulet2021}. Different values for the tidal $Q_p$ are correlated with present-day ages in the solution space: having an age over a gigayear \citep[which seems likely for this system; ][]{Yoshida2023} would indicate that the planetary $Q_p$ is probably not much less than $10^6$, lest the planet have long since finished circularizing.

Our calculations demonstrate that tidal inflation is a plausible way to explain the planet's large radius. However, they also demonstrate the strong degeneracies inherent to the problem: all available data for TOI-1420b can be matched equally well in models where the planet is young and circularizing or old and very gas-rich. With relatively uninformative prior information on TOI-1420b's eccentricity ($<0.17$) and age \citep[$<10.7$~Gyr,][]{Yoshida2023}, these degeneracies cannot be broken. Future efforts to more precisely constrain the eccentricity and/or age of the planet would therefore be highly valuable for understanding TOI-1420b's evolutionary state.

\subsection{Can high-altitude hazes ``deflate'' TOI-1420b?}

The flat transmission spectra of lower-mass super-puffs are well-modeled by high-altitude aerosols, which may also be responsible for their large apparent radii \citep{Kawashima2019, Gao2020, Ohno2021}. It is worth considering this hypothesis in more detail for the higher-mass super-puff TOI-1420b, as the model requires strong outflows (similar to the one we have detected) to carry aerosols to high altitudes. While high-altitude aerosols are ruled out as a possibility for WASP-107b \citep[molecular features are detected lower in its atmosphere;][]{Spake2018, Kreidberg2018, Dyrek2023}, transmission spectroscopy sensitive to the lower atmosphere of TOI-1420b is not yet available.

Hazes can persist at $\mu$bar to nbar pressures in atmospheres undergoing outflows \citep{Wang2019,Gao2020,Ohno2021}, and so the 1~nbar radius should represent the maximum extent to which hazes can increase a planet's radius from a clear atmosphere case. As such, to test whether TOI-1420b could be a smaller planet with optically thick high-altitude hazes, we compute the 1~nbar radius of this planet using the atmospheric and planetary structure model of \citet{Gao2020}. Briefly, the model assumes a core with Earth-like composition and a mass--radius relation adopted from \citet{zeng2019}, which lies beneath a gas envelope of solar composition and a simplified thermal structure consisting of a deep convective zone and an upper radiative zone that is connected to the convective zone at the radiative-convective boundary (RCB). The convective zone structure is solved following \citet{owen2017} assuming hydrostatic equilibrium and a polytrope equation of state with a power of 7/5 to represent a largely H$_2$ composition. At the RCB, the temperature-pressure profile transitions smoothly to a radiative profile following \citet{piso2014}, with the temperature of the RCB set to the equilibrium temperature of TOI-1420b. For the opacity in the radiative zone we use the tables from \citet{freedman2014}. We refer the reader to \citet{Gao2020} for more details regarding the model. 

\begin{figure}[t]
    \centering
    \includegraphics[width=0.5\textwidth]{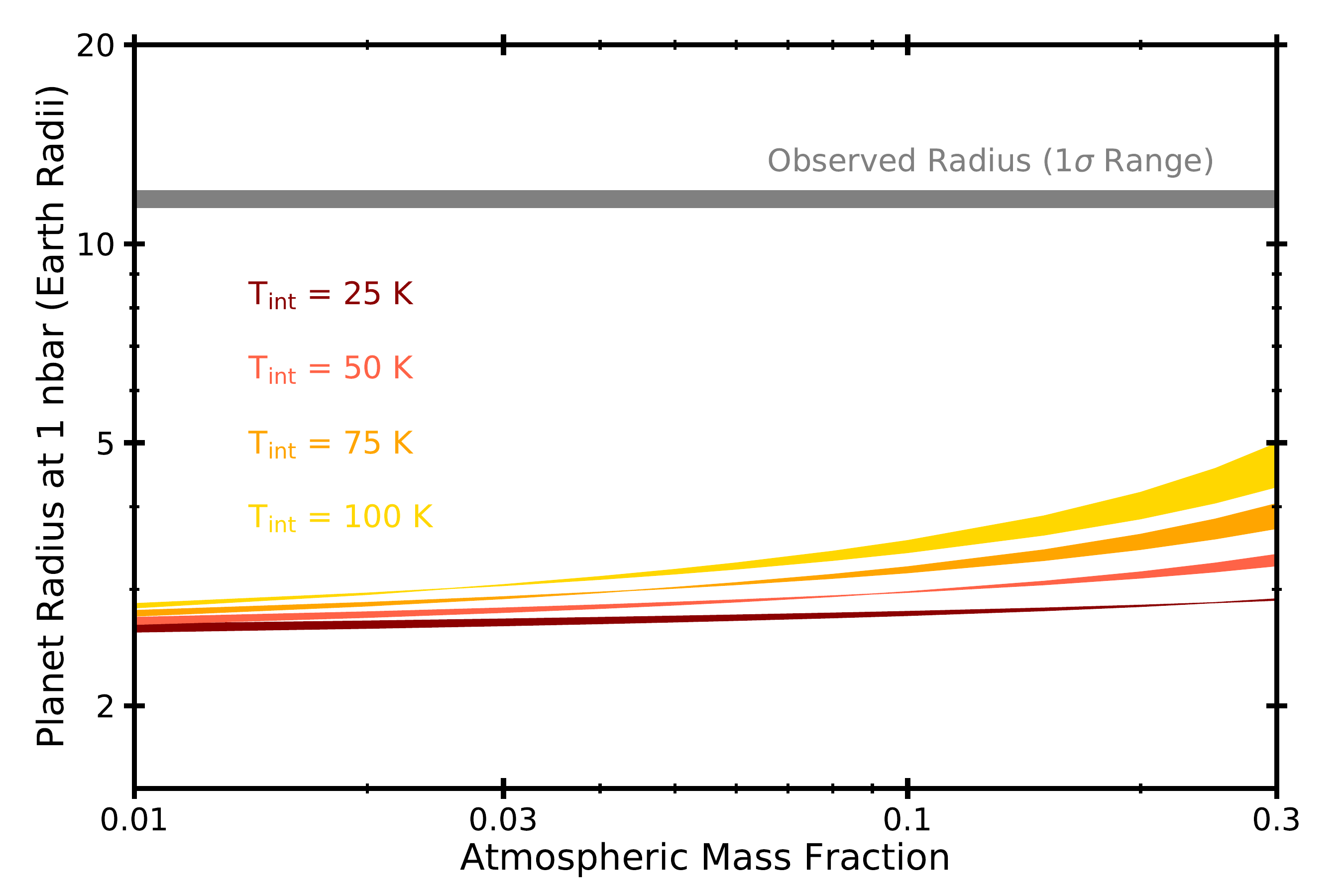}
    \caption{Predicted planet radii at a pressure level of 1~nbar in the atmosphere for a range of atmospheric mass fractions and intrinsic temperatures T$_{\rm int}$ of 25~K (dark red), 50~K (red), 75~K (orange), and 100~K (yellow). The span in the shaded regions represents the impact of the uncertainty in the planet mass on the model. The 1$\sigma$ range of the observed radius is shown in the gray region. For atmospheric mass fractions $<$0.3, no model is able to match the observed radius for a haze that is optically thick in the slant geometry at 1~nbar.}
    \label{fig:hazes}
\end{figure}

In Figure~\ref{fig:hazes}, we show that the 1~nbar radius of TOI-1420b does not reach the observed radius for any atmospheric mass fractions $<$30\% and intrinsic temperatures $<$100 K, even when the uncertainty in the mass is taken into account. These limits roughly correspond to those of Neptune-mass planets with the age and insolation of TOI-1420b \citep{Helled2011, lopez2014}. Therefore, it is unlikely TOI-1420b is a smaller planet that has been made to look larger with a high-altitude haze layer. 

\section{Conclusion}
\label{sec:conc}
Many lower-mass super-puffs ($M_p < 10M_\Earth$) are expected to have atmospheric lifetimes shorter than their ages, while higher-mass ($M_p > 10M_\Earth$) super-puffs are expected to be stable to catastrophic atmospheric loss even when strongly irradiated. To test this expectation, we searched for an outflow in the atmosphere of TOI-1420b, a warm ($T_\mathrm{eq} = 960$~K), higher-mass ($M_p = 25.1M_\Earth$) super-puff with a density of just $\rho = 0.08$~g~cm$^{-3}$ \citep{Yoshida2023}. We observed one full transit and one partial transit with the narrowband helium filter on Palomar/WIRC and compared the resulting light curve to the TESS light curve. We found that the transit in the helium bandpass was consistently $0.671\pm0.079\%$ deeper than the TESS transit, indicating excess absorption by outflowing metastable helium at a confidence of 8.5$\sigma$.

We first interpreted our observations using the 1D photoionization hydrodynamics code \textsf{ATES} coupled to \textsf{Cloudy}, which allowed for a self-consistent treatment of the outflow thermodynamics. The model that best matched our data used a low activity proxy for the host star's XUV spectrum ($1/3\times$ solar XUV) with a hydrodynamic escape parameter $\lambda_p \approx 5$. Using this XUV spectrum and escape parameter, we then calculated a three-dimensional isothermal outflow model using \textsf{Athena++}. This model gave a good match to the observed excess absorption at mid-transit with a mass-loss rate of $\dot{M} = 4.9\times10^{10}$~g~s$^{-1}$. The corresponding mass-loss timescale is $M_p/\dot{M} \approx 70$~Gyr; despite its remarkably low density, the mass-loss rate of TOI-1420b is quite modest. The situation is similar for the other warm high-mass super-puff WASP-107b, which has $M_p/\dot{M} \approx 30$~Gyr \citep{Wang2021}. Our observations confirm that super-puffs with $M_p > 10M_\Earth$ are easily able to retain their massive atmospheres over long timescales. These planets may therefore have formed with unusually massive envelopes, likely accreting gas outside the water ice line before migrating to their present positions \citep{Stevenson1984, Venturini2015, Lee2016, Piaulet2021}. 

We also explored a number of alternative hypotheses for TOI-1420b's large radius. We presented some illustrative evolution calculations showing that all available data for TOI-1420b could be matched at a larger bulk metallicity \citep[more consistent with models of core accretion in the inner disk;][]{Lee2019} provided the planet recently underwent tidal circularization. On the other hand, we found that high-altitude hazes are disfavored as an explanation for TOI-1420b's large radius. The planet is simply too massive, resulting in a relatively small scale height among the super-puffs: even optically-thick hazes at nbar~pressure for a Neptune-like planet would fall well short of explaining TOI-1420b's large radius. This bodes well for future atmospheric characterization efforts: we do not anticipate that atmospheric features will be completely obscured by hazes, similar to the case for WASP-107b \citep{Kreidberg2018, Spake2018, Dyrek2023}.

TOI-1420b's helium absorption is the largest signal we have observed at high confidence with Palomar/WIRC thus far \citep{Vissapragada2022:helium}. We encourage other investigators to follow up the helium absorption for this planet with high-resolution spectroscopy, which would provide additional constraints on the atmospheric modeling. With such a strong signal this should also be a good candidate to search for a Doppler-shifted line core and/or extended egress, which is indicative of an outflow morphology sculpted by stellar winds in WASP-69b and WASP-107b \citep{Nortmann2018, Spake2021, Wang2021, MacLeod2022, Tyler2024}. Our light curve of TOI-1420b appears symmetric, and is thus well-fit by a 3D model where the stellar wind is not strong enough to shepherd the outflowing material into a comet-like tail, but tail features can be challenging to capture at low effective resolving power from the ground \citep{Vissapragada2020:helium, Paragas2021, Fu2022}. High-resolution studies of TOI-1420b would therefore enable more direct morphological comparisons to planets like WASP-107b, and may even enable comparative studies of the stellar wind strengths in these systems.

Lower-mass super-puffs are expected to be even more susceptible to atmospheric escape than TOI-1420b and WASP-107b. In particular, there are a number of planets in Figure~\ref{fig:superpuff} that are similar in equilibrium temperature to WASP-107b, including TOI-216b, Kepler-79d, Kepler-18d, and Kepler-223e. If their host stars are active, these planets might also exhibit triplet helium absorption. While many super-puffs reside in systems that are too faint for current facilities, a number of suitable targets have been discovered in bright systems, including K2-24c, HIP 41378 f, and TOI-216b \citep{Petigura2018, Santerne2019, McKee2023}. Although the transit durations for these planets tend to be long, helium absorption could still be constrained with multi-night ground-based campaigns \citep{Gully-Santiago2023, Zhang2023} or space-based observations with \textit{JWST} \citep{Fu2022, DosSantos2023}. By characterizing the upper atmospheres of lower-mass super-puffs, we can gain greater insight into the nature of these remarkably low-density worlds.

\begin{acknowledgments}

We thank Paul Nied and Diana Roderick for assistance with telescope operations. We additionally thank Fei Dai, Ryan Rubenzahl, and Jeremy Drake for helpful conversations. This work was supported by telescope time allocated to NASA-NSF Exoplanet Observational Research (NN-EXPLORE) partnership through the scientific partnership of the National Aeronautics and Space Administration, the National Science Foundation, and the NOIRLab. This research made use of \textsf{photutils}, an \textsf{astropy} package for detection and photometry of astronomical sources \citep{photutils}. This research made use of \textsf{exoplanet} \citep{exoplanet:joss,
exoplanet:zenodo} and its dependencies \citep{exoplanet:agol20,
exoplanet:arviz, exoplanet:astropy13, exoplanet:astropy18, exoplanet:kipping13,
exoplanet:luger18, exoplanet:pymc3, exoplanet:theano}.

\facilities{ADS, NASA Exoplanet Archive, Hale 200-inch (WIRC)}
\software{\textsf{photutils} \citep{photutils}, \textsf{exoplanet} \citep{exoplanet:joss,
exoplanet:zenodo}, \textsf{pymc3} \citep{exoplanet:pymc3}, \textsf{arviz} \citep{exoplanet:arviz}, \textsf{numpy} \citep{numpy}, \textsf{scipy} \citep{scipy}, \textsf{astropy} \citep{exoplanet:astropy13, exoplanet:astropy18, AstropyCollaboration2022}, \textsf{p-winds} \citep{DosSantos2022:pwinds}, \textsf{ATES} \citep{Caldiroli2021}, \textsf{Cloudy} \citep{Ferland1998, Ferland2017}, \textsf{Athena++} \citep{Stone2008, Stone2020}}

\end{acknowledgments}

\clearpage
\appendix

\section{Posterior Probability Distributions}
\begin{figure}[b]
    \centering
    \includegraphics[width=0.8\textwidth]{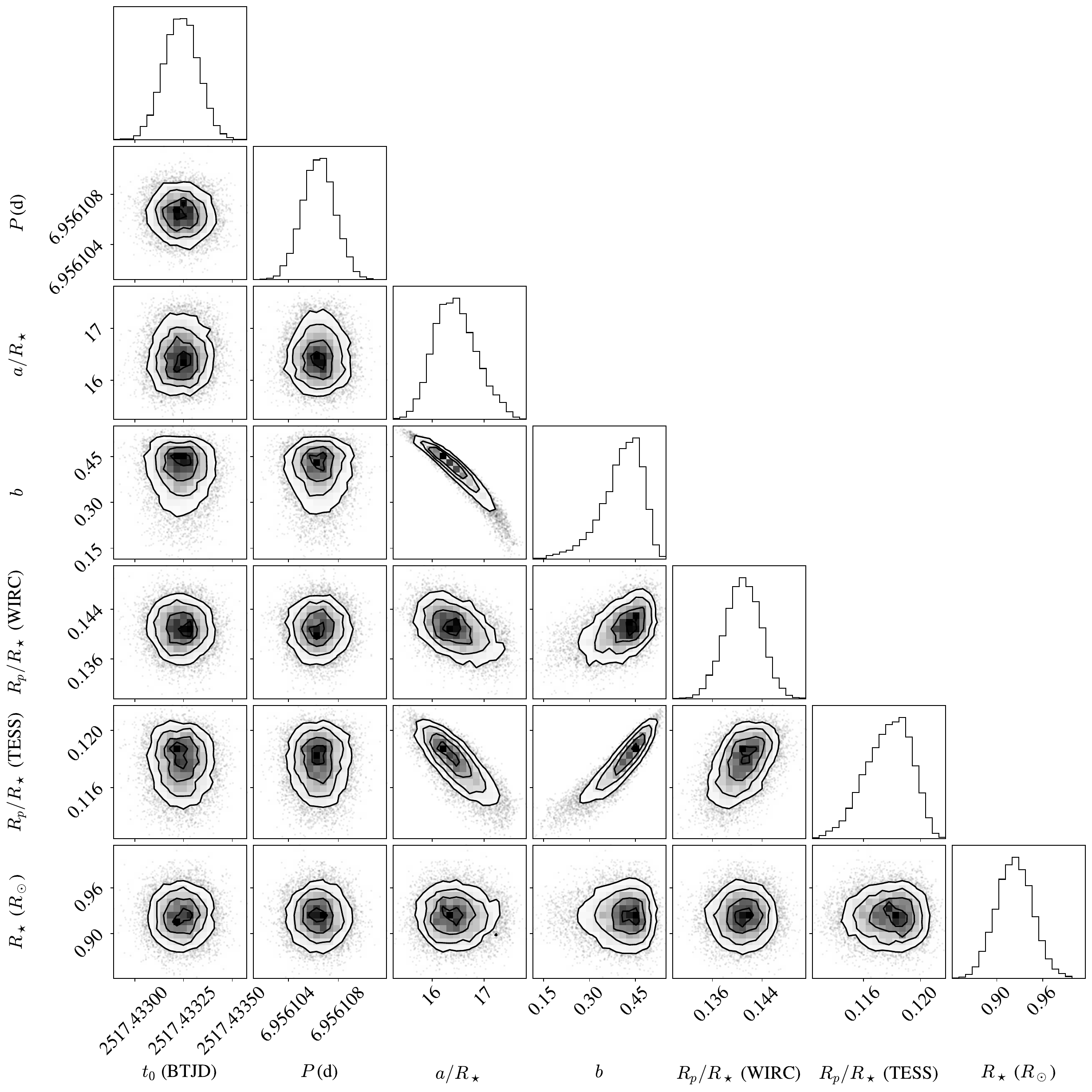}
    \caption{Corner plot for key fitting parameters in the final joint fit.}
    \label{fig:corner}
\end{figure}

Here, we provide a corner plot for the key fitting parameters of the final joint fit in Figure~\ref{fig:corner} and a table summarizing the posteriors for detrending weights and error scaling parameters in Table~\ref{table:extraposteriors}.

\begin{deluxetable}{ccccc}
\tablecaption{Priors and posteriors for the TESS error scaling parameter, jitters, and detrending weights. \label{table:extraposteriors}}
\tablehead{\colhead{Parameter} & \colhead{Prior} & \colhead{Night 1} & \colhead{Night 2} & \colhead{Night 1 + Night 2}}
\startdata
$k_\mathrm{TESS}$ & $\mathcal{U}(0.5, 1.5)$ & $0.9158_{-0.0022}^{+0.0021}$ & $0.9158_{-0.0022}^{+0.0022}$ & $0.9158_{-0.0022}^{+0.0021}$\\
$\sigma_\mathrm{WIRC, 1}$ & $\mathcal{U}(10^{-6}, 10^{-2})$ & $0.00204_{-0.00067}^{+0.00053}$ & -- & $0.00202_{-0.00065}^{+0.00053}$\\
$w_{1,1}$ &  $\mathcal{U}(-2, 2)$ & $0.073_{-0.035}^{+0.034}$ & -- & $0.070_{-0.034}^{+0.035}$\\
$w_{2,1}$ & $\mathcal{U}(-2, 2)$ & $0.719_{-0.078}^{+0.079}$ &  -- & $0.719_{-0.075}^{+0.076}$\\
$w_{3,1}$  & $\mathcal{U}(-2, 2)$ & $0.0072_{-0.0393}^{+0.0402}$ & -- & $0.0083_{-0.0414}^{+0.0410}$\\
$w_{4,1}$ & $\mathcal{U}(-2, 2)$ & $0.014_{-0.036}^{+0.037}$ & -- &
$0.012_{-0.036}^{+0.037}$\\
$w_{5,1}$  & $\mathcal{U}(-2, 2)$ & $0.129_{-0.051}^{+0.053}$ & -- & $0.138_{-0.051}^{+0.049}$\\
$w_{6,1}$ & $\mathcal{U}(-2, 2)$ & $0.072_{-0.038}^{+0.038}$  & -- & $0.068_{-0.039}^{+0.039}$\\
$w_{7,1}$ &  $\mathcal{U}(-2, 2)$  & $0.0146_{-0.0052}^{+0.0051}$ & -- & $0.0152_{-0.0052}^{+0.0050}$\\
$\sigma_\mathrm{WIRC, 2}$ & $\mathcal{U}(10^{-6}, 10^{-2})$ & -- & $0.00361_{-0.00049}^{+0.00051}$ & $0.00361_{-0.00049}^{+0.00050}$\\
$w_{1,2}$ & $\mathcal{U}(-2, 2)$  & -- & $0.254_{-0.092}^{+0.095}$ & $0.250_{-0.094}^{+0.093}$\\
$w_{2,2}$ &  $\mathcal{U}(-2, 2)$  & -- & $0.44_{-0.12}^{+0.13}$ & $0.43_{-0.12}^{+0.12}$\\
$w_{3,2}$ & $\mathcal{U}(-2, 2)$  & -- & $0.093_{-0.047}^{+0.046}$ & $0.098_{-0.045}^{+0.044}$\\
$w_{4,2}$ &  $\mathcal{U}(-2, 2)$  & -- & $0.083_{-0.050}^{+0.050}$ &$0.093_{-0.051}^{+0.050}$ \\
$w_{5,2}$ & $\mathcal{U}(-2, 2)$  & -- & $0.034_{-0.090}^{+0.087}$ & $0.027_{-0.087}^{+0.090}$\\
$w_{6,2}$ & $\mathcal{U}(-2, 2)$  & -- & $0.121_{-0.060}^{+0.060}$ &  $0.116_{-0.061}^{+0.062}$\\
$w_{7,2}$ & $\mathcal{U}(-2, 2)$ & -- & $-0.006_{-0.016}^{+0.016}$ &  $0.007_{-0.013}^{+0.013}$\\
\enddata
\end{deluxetable}

\clearpage
\bibliography{references}

\end{document}